\documentclass[slac_one]{revtex4}
\usepackage{graphicx}
\usepackage{fancyhdr}
\begin{document}

\title{{\small{2005 ALCPG \& ILC Workshops - Snowmass,
U.S.A.}}\\ 
\vspace{12pt}
New Developments in Extra-dimensional Dark Matter} 

%

\author{Jose A.~R.~Cembranos}
\affiliation{Department of Physics and Astronomy,
 University of California, Irvine, CA 92697 USA}

\author{Antonio Dobado}
\affiliation{Departamento de F\'{\i}sica Te\'orica I,
Universidad Complutense de Madrid, 28040 Madrid, Spain}

\author{Jonathan L.~Feng}
\affiliation{Department of Physics and Astronomy,
 University of California, Irvine, CA 92697 USA}

\author{Antonio L.~Maroto}
\affiliation{Departamento de F\'{\i}sica Te\'orica I,
Universidad Complutense de Madrid, 28040 Madrid, Spain}
\author{Arvind Rajaraman}
\affiliation{Department of Physics and Astronomy,
 University of California, Irvine, CA 92697 USA}
\author{Fumihiro Takayama}
\affiliation{Institute for High-Energy Phenomenology,
Cornell University, Ithaca, NY 14853, USA}

\begin{abstract}
We summarize the main features of several dark matter candidates in
extra-dimensional theories. In particular, we review Kaluza-Klein (KK)
gravitons in universal extra dimensions and branons in brane-world
models. KK gravitons are superWIMP (superweakly-interacting massive
particle) dark matter, and branons are WIMP (weakly-interacting
massive particle) dark matter.  Both dark matter candidates are
naturally produced in the correct amount to form much or all of dark
matter.
\end{abstract}

\maketitle

\thispagestyle{fancy}


\section{KALUZA-KLEIN GRAVITONS}

\begin{figure}
\centering
\resizebox{8cm}{!}{\includegraphics{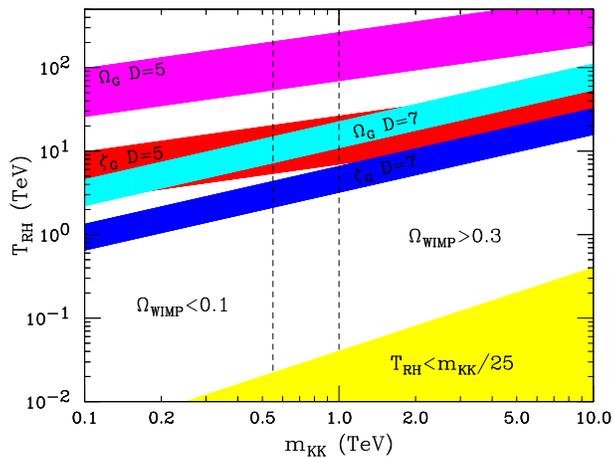}}
\caption{Overclosure and BBN constraints on the reheat temperature
$T_{RH}$, as a function of the Kaluza Klein mass $m_{KK}$.  The
vertical bands delimit regions of $B^1$ thermal relic abundance
$\Omega_{WIMP}$. See~\cite{Feng:2003nr} for
details. \label{fig:reheat}}
\end{figure}

In models with universal extra dimensions
(UED)~\cite{Appelquist:2000nn}, gravity and all Standard Model (SM)
fields have access to the entire space, which is called bulk space.
The first studies of such models concentrated on the Kaluza-Klein (KK)
partners of SM particles, but all UED models necessarily also have KK
gravitons, and these may be viable DM candidates. Given the general
formalism for analyzing the dynamics of gravitons in UED
theories~\cite{Feng:2003nr}, one can find the widths for decays of KK
fermions and KK gauge bosons into KK gravitons.  These results are of
special relevance when a KK graviton is the lightest KK particle and a
superWIMP candidate~\cite{SWIMP}, as they determine the observable
implications of KK graviton DM for Big Bang Nucleosynthesis (BBN)
analyses, the cosmic microwave background, the diffuse photon
flux~\cite{Feng:2003nr} and structure formation~\cite{CFRT}.  The
possibility of populating a large number of graviton states at
different KK levels implies that the production of gravitons after
reheating is extremely efficient and extremely sensitive to the reheat
temperature $T_{RH}$. The constraints on $T_{RH}$ are therefore
stringent (see Figure~\ref{fig:reheat}).

This superWIMP scenario can be studied in colliders. Decays to KK
gravitons may be observed by trapping KK lepton next lightest KK
particles in water tanks placed just outside collider detectors.  By
draining these tanks periodically to underground reservoirs, slepton
decays may be observed in quiet environments as in the gravitino
case~\cite{Buchmuller:2004rq,Feng:2004gn,Hamaguchi:2004df,Feng:2004yi,Brandenburg:2005he,Feng:2005gj}.
Precision studies of KK lepton decays are therefore possible and can
provide direct observations of gravitational effects at colliders;
measurements of the extra-dimensional size and Newton's constant;
precise determinations of the KK graviton's contribution to DM and
laboratory studies of Big Bang nucleosynthesis and cosmic microwave
background phenomena.

\section{BRANONS}

The brane-world (BW) scenario is defined by the fact that the SM particles are
restricted to a three-dimensional hypersurface or 3-brane, whereas the
gravitons can propagate along the whole bulk space.  In flexible BW
models (when the brane tension scale $f$ is much
smaller than the $D$ dimensional or fundamental gravitational scale
$M_D$, i.e. $f<<M_D$), the branons are the only new relevant
low-energy particles~\cite{DoMa}. The SM-branon low-energy effective
Lagrangian reads~\cite{DoMa,BSky,ACDM}:
\begin{eqnarray}
{\mathcal L}_{Br}&=& \frac{1}{2}g^{\mu\nu}\partial_{\mu}\pi^\alpha
\partial_{\nu}\pi^\alpha-\frac{1}{2}M^2\pi^\alpha\pi^\alpha
\frac{1}{8f^4}(4\partial_{\mu}\pi^\alpha
\partial_{\nu}\pi^\alpha-M^2\pi^\alpha\pi^\alpha g_{\mu\nu})
T^{\mu\nu}_{SM}\label{lag}\,.
\end{eqnarray}
Branons interact by pairs with the SM energy-momentum tensor and their
couplings are suppressed by the brane tension $f^4$. In fact, they are
generically stable and weakly interacting. These features make them
natural DM~\cite{CDM,M} candidates. On the other hand, the branon
phenomenology is very rich since they are coupled with the entire
SM. The branon signals can be characterized by their number $N$, the
brane tension scale $f$, and their masses $M$. The collider signatures
for direct branon production are missing energies. In particular, the
monojet and single photon signals for hadron colliders~\cite{ACDM} and
the single Z and single photon channels for $e^+e^-$
colliders~\cite{ACDM}. This last signature has been studied
experimentally by the L3 Coll.~\cite{L3} finding the most constraining
bound for the brane tension scale: $f>180$ GeV, for light
branons~\cite{ACDM,L3,CrSt}). On the other hand, branon radiative
correction~\cite{rad} modify four body interactions, electroweak
precision observables, anomalous magnetic moments and Higgs boson
phenomenology.

\begin{figure}[h]
\centering
\resizebox{9cm}{!}{\includegraphics{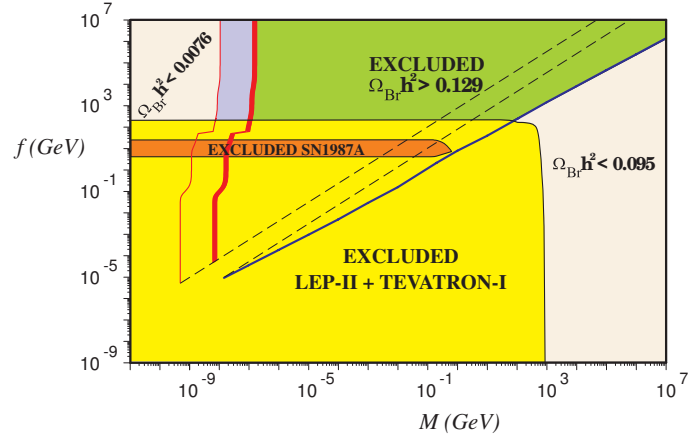}} \caption {Relic
abundance in the $f-M$ plane for a model with one branon of mass
$M$. The two lines on the left correspond to the
$\Omega_{Br}h^2=0.0076$ and $\Omega_{Br}h^2=0.129 - 0.095$ curves for
hot-warm relics, whereas the right line corresponds to the latter
limits for cold relics (see~\cite{CDM} for details).  The lower area
is excluded by single-photon processes at LEP-II \cite{ACDM,L3}
together with monojet signal at Tevatron-I \cite{ACDM}. The
astrophysical constraints are less restrictive and they mainly come
from supernova cooling by branon emission \cite{CDM}.} \label{mother}
\end{figure}

\section{Conclusions}

In this work we have reviewed the main features of KK gravitons in UED
and branons in brane-worlds.  The thermal abundance of branons, and the
non-thermal abundance of KK gravitons from late decays, are both
naturally in the right range to form a significant component of dark
matter.  We have considered the main phenomenological signals and
constraints of these dark matter scenarios.

\begin{acknowledgments}

JARC acknowledges the hospitality and collaboration of workshop
organizers and conveners, and economical support from the NSF and
Fulbright OLP.
The work of JARC is supported in part by NSF grant No.~PHY--0239817,
the Fulbright-MEC program, and the FPA 2005-02327 project (DGICYT,
Spain).  The work of AD is supported in part FPA 2004-02602 and FPA
2005-02327 project (DGICYT, Spain).  The work of JLF is supported in
part by NSF grant No.~PHY-0239817, NASA grant No.~NNG05GG44G, and the
Alfred P.~Sloan Foundation.  The work of ALM is supported in part FPA
2004-02602 and FPA 2005-02327 project (DGICYT, Spain).  The work of AR
is supported in part by NSF grant No.~PHY-0354993.  The work of FT is
supported by the NSF grant No.~PHY-0355005.

\end{acknowledgments}

\end{document}